\documentclass[letter,scriptaddress,twocolumn, prl,showkeys]{revtex4}
\usepackage{eurosym}
\usepackage{amsfonts}
\usepackage{amsmath}
\usepackage{amssymb}
\usepackage{graphicx}
\usepackage[colorlinks,hyperindex]{hyperref}
\usepackage{footnote}

\hypersetup{
colorlinks,
citecolor=blue,
linkcolor=black,
urlcolor=black,
}
\DeclareMathOperator{\sech}{sech}
\providecommand{\U}[1]{\protect\rule{.1in}{.1in}}

\begin{document}

\title[\textcolor{black}{Oscillons in $\phi^6$-theories: Possible occurrence in MHD}]{\textcolor{black}{Oscillons in $\phi^6$-theories: Possible occurrence in MHD}}
\author{R. A. C. Correa$^{\textcolor{red}{1,2}}$ 
\footnote{E-mail: \textcolor{blue}{fis04132@gmail.com}}, 
W. de Paula$^{\textcolor{red}{2}}$ 
\footnote{E-mail: \textcolor{blue}{wayne@ita.br}}, 
T. Frederico$^{\textcolor{red}{2}}$ 
\footnote{E-mail: \textcolor{blue}{tobias@ita.br}}, 
O. Oliveira$^{\textcolor{red}{3}}$ 
\footnote{E-mail: \textcolor{blue}{orlando@uc.pt}}, and 
F. E. M. Silveira$^{\textcolor{red}{4}}$ 
\footnote{E-mail: \textcolor{blue}{francisco.silveira@ufabc.edu.br}}}
\affiliation{$^{\textcolor{red}{1}}$SISSA - International School for Advanced Studies,
Via Bonomea 265, I-34136, Trieste, Italy,}
\affiliation{$^{\textcolor{red}{2}}$Instituto Tecnol\'{o}gico de Aeron\'{a}utica, DCTA,
12228-900, S\~{a}o Jos\'{e} dos Campos, SP, Brazil,}
\affiliation{$^{\textcolor{red}{3}}$CFisUC, Department of Physics, University of Coimbra,
P-3004 516 Coimbra, Portugal,}
\affiliation{$^{\textcolor{red}{4}}$CCNH, Universidade Federal do ABC, 09210-170, Santo
Andr\'{e}, SP, Brazil}
\pacs{PACS number}

\begin{abstract}
In this work, we report on the possibility of occurrence of oscillon 
configurations in the fourth state of matter. Oscillons are extremely 
long-lived, time-periodic, spatially-localised scalar field structures. 
Starting from a scalar field theory in 1+1 space-time dimensions, we 
find out that small-amplitude oscillons can be obtained in the 
framework of a $\phi^6$ self-interacting potential. A connection 
between our results and ideal MHD theory is established. Perspectives 
for a development of the present work are pointed out.
\end{abstract}

\maketitle

\section{Introduction}

It is widely acknowledged that, under certain circumstances, a large number
of natural systems can exhibit a nonlinear behavior. For instance, we may
observe nonlinearities in the condensed state of matter 
\cite{nature-417-2002}, elementary particle physics \cite{weinberg}, 
cosmological scenarios \cite{PRL-95-2005}, biological systems 
\cite{nature-9-2018}, but also in plasma physics 
\cite{PRL-85-2000,nature-431-2004}.

In classical field theories, there is an important class of configurations,
termed solitons \cite{rajaraman}, whose existence is entirely due to the
nonlinearity of the field equations. Such solutions are well known in
Lorentz and CPT breaking systems \cite{rafaelPRD2011}, modified theories of
gravitation \cite{rafaelEPJC2016}, two-dimensional N=1 supersymmetric
quantum field theories \cite{mussardoJHEP2007}, non-integrable quantum field
theories \cite{mussardoNPB2007}, fibre optics \cite{nature-6-2010}, as well
as in plasma physics \cite{lonngrenPP1983,bailungPRL2011}. A soliton is a
static field configuration whose energy density profile is localised in
space. In particular, it exhibits the distinctive property of to retain its
shape after collision with another soliton.

Quite interestingly, in the 1970s, a new class of localised nonperturbative
solutions, which may be derived in the realm of nonlinear theories, was
reported \cite{dashenPRD1975, KudryJETP1975, BogoJETP1977, MakPR1978}. Their
most noticeable features were the time-dependence, extreme long-life, and
spatial localisation. In spite of such an intriguing behavior, only in the
1990s their analytical properties were firstly explored by Gleiser 
\cite{gleiserPRD1994}, who coined them oscillons \cite{gleiserPRD1995}.

Oscillons have attracted attention from several areas of research, namely
standard model-extensions \cite{rafaelPRD2015}, supersymmetry theories 
\cite{rafaelPLB2018}, gravitational waves \cite{stefanJCAP2018,jingPRD2018}, 
high energy systems in presence of external potentials \cite{romanJHEP2018}, 
the Abelian-Higgs model \cite{diakonosPRE2015, achilleosPRD2013}, certain
Lorentz violating scenarios \cite{rafaelAHEP2015}, spontaneous symmetry
breaking phenomena \cite{gleiserPRD2012}, and cosmological background
investigations \cite{aminPRD2014-2018}. Such an interest may be attributed
to the unexpected longevity combined with nearly periodic oscillation in
time, as exhibited by those structures.

Nonlinearities may be also observed in the fourth state of matter. Actually,
solitons are ubiquitous in ivestigations of nonlinear dispersive media 
\cite{zaburkyPRL1965}, electron-beam plasmas \cite{berthPP2000}, weakly
relativistic plasmas containing electrons, positrons, and ions 
\cite{gillPLA2007}, fermionic quantum plasmas \cite{marklundPRE2007}, dusty
plasmas \cite{shuklaNJP2003, singhPP2018}, plasma slabs 
\cite{stenfloPRE1995}, and cold plasma columns bounded by deformable dielectrics 
\cite{stenfloPP1995}. Recently, a novel regime of soliton-plasma interaction has
been found in a gas-filled hollow-core photonic crystal fiber 
\cite{taniPRL2017}. Despite the impressive list, as mentioned above, matching
solitons to plasmas, investigations linking oscillons to the latter from
first principles still lack.

Given that oscillons are known to play a central role in the nonlinear
dynamics of a wide number of physical systems, in this work, we propose to
explore their possible emergence on a plasma background. Thus, we shall
consider a scalar field theory with nonlinear interactions responsible to
preserve the localisation of energy for a remarkably long time. In
particular, we shall regard only small amplitude oscillons living in 1+1
space-time dimensions \cite{fodorPRD2006, fodorPRD2008}. The reason to
confine ourselves to this special regime stems from the flexibility it
offers to apply our results to a definite plasma system. However, as we will
show, our approach opens up a new window to explore a large number of
nonlinear plasma scenarios.

\bigskip

\section{Scalar Field Dynamics}

Let us start by considering a real scalar field theory (SFT) in 1+1
space-time dimensions, described by the action

\begin{equation}
S_{\mathrm{SFT}}=\int dt \int dx \left[\frac{(\partial_{t}\phi)^{2}}{2} 
-\frac{(\partial_{x}\phi)^{2}}{2}-U\right] ,  \label{1}
\end{equation}

\noindent where we abbreviate $\partial _{t}=\partial /\partial t$ and 
$\partial _{x}=\partial /\partial x$. The quantity $U=U(\phi; \partial
_{t}\phi, \partial _{x}\phi)$ is a potential functional, with 
$\phi=\phi(t;x) $ denoting the scalar field. We remark that $U$ depends on
both $\phi$ and its derivatives. Our motivation to consider that instance
stems from the observation that nonlinear plasma dynamics usually exhibits
such a behavior. \textcolor{black}{Actually, the natural framework at which oscillon configurations are most likely to be analy\-tically described in plasmas is the so-called reduced magnetohydrodynamics (RMHD). That formulation allows the identification of three independent time scales, which may be associated with (1) MHD equilibrium, (2) perturbations normal to the externally applied magnetic field $\vec B$, and (3) those parallel to $\vec B$ \cite{krugerPoP1998}. A suitable account of (1), by an appropriate choice of geometry, and elimination of (2), by an adequate requirement of cons\-traints, enables a satisfactory description of (3). In fact, such an approach has been proved consistent with the usual assumption of a perfectly conducting fluid (the so-called ideal limit) in magnetohydrodynamics, even at sufficiently high perturbative frequencies (higher than the inverse Alfv\'en transit time scale), by robust numerical treatments \cite{oughtonPoP2004}. In 2+1 dimensions, RMHD may be derived from a variational principle, whose action depends on both the fields and their derivatives \cite{brizardPoP2010}. A somewhat more general derivation is possible in the realm of kinetic theory \cite{burbyJPP2016}. In this work, we obtain the aforementioned action dependence, in 3+1 dimensions, through a simple perturbative procedure.}

On general grounds, let us fix the potential,

\begin{eqnarray}
U=V &+&a\phi \partial _{t}\phi +b\phi \partial _{x}\phi +c\left( \partial
_{t}\phi \right) ^{2} +d\left( \partial _{x}\phi \right) ^{2}  \notag \\
&+&f\phi ^{F}\partial _{t}\phi +g\phi ^{G}\partial _{x}\phi +h\phi
^{H}\partial _{t}\phi \partial _{x}\phi ,  \label{2}
\end{eqnarray}

\noindent where the coefficients $a$, $b$, $c$, $d$, $f$, $g$, and $h$, and powers $F$, $G$, and $H$ are constant. \textcolor{black}{We notice that $a$, $b$, $f$, and $g$ have dimension of mass, while $c$, $d$, and $h$ are dimensionless}. The functional $V=V(\phi )$ is a particular self-interacting potential (see below). \textcolor{black}{Before we proceed, it is important to remark that since we are seeking confi\-gurations that exhibit non-trivial topology, for which the scalar field does not go to zero for infinite time asymptotic values, the terms proportional to the coefficients $a$, $b$, $f$, and $g$ cannot be written as surface terms in the action. In addition, we observe that by assuming $c\neq d$, the action does not remain Lorentz invariant. Interestingly, that would allow for the investigation of the impact of Lorentz violation in such a context. However, that ins\-tance is beyond the scope of this paper.} 

An arbitrary variation of the action (\ref{1}) leads to the classical field
equation

\begin{eqnarray}
V_{\phi }&+&(1+2c)\partial _{tt}\phi-(1-2d)\partial _{xx}\phi  \notag \\
&-&Hh\phi ^{H-1}\partial _{t}\phi \partial _{x}\phi -2h\phi ^{H}\partial
_{tx}\phi=0,  \label{3}
\end{eqnarray}

\noindent where use has been made of Eq. (\ref{2}), and we have abbreviated 
$V_{\phi}=dV/d\phi$, and $\partial_{tt}=\partial^2/\partial t^2$, 
$\partial_{xx}=\partial^2/\partial x^2$, and $\partial_{tx} 
=\partial^2/\partial t\partial x$.

In the sequel, it proves useful to consider the dilations $t=T(1+2c)^{1/2}$
and $x=X(1-2d)^{1/2}$. Thus, Eq. (\ref{3}) reads as

\begin{eqnarray}
V_{\phi}&+&\partial _{TT}\phi-\partial _{XX}\phi -\frac{Hh\phi
^{H-1}\partial _{T}\phi \partial _{X}\phi } {(1+2c)^{1/2}(1-2d)^{1/2}} 
\notag \\
&-&\frac{2h\phi ^{H}\partial _{TX}\phi } {(1+2c)^{1/2}(1-2d)^{1/2}}=0.
\label{4}
\end{eqnarray}

\noindent Of course, now $\phi =\phi (T;X)$.

In this work, we confine ourselves to the consideration of the well-known
self-interacting potential $\phi^{6}$, which plays a central role in the
description of first-order phase transitions \cite{mussardoBOOK} and particle physics
phenomenology \cite{weinberg}. Such a symmetric model has been also a popular point of
depart for several investigations concerning oscillon dynamics \cite{fodorPLB2009,fodorPRD2009,rafaelAHEP2016}. 
\textcolor{black}{In particular, we consider the $\phi^{6}$ potential, motivated by investigations on oscillons, 
in an expanding universe \cite{aminPRD2010}.} We write

\begin{equation}
V(\phi)=\frac{\omega^2\phi^2}{2} \left(1-\frac{2\phi^2}{3\phi_0^2}\right)^2,
\label{5}
\end{equation}

\noindent where $\omega$ and $\phi_0$ are real, positive valued parameters,
and the assigned prefactors have been suitably chosen. \textcolor{black}{The quantity $\omega$ has dimension of mass. In the context of the standard model, one may add the $\phi^{6}$ interaction to the Higgs potential, in order to describe the strong first-order electroweak phase transition for masses, thereby leading to scales above $100$ GeV \cite{gevJHEP2005}.}

The profile of $V(\phi)$ is illustrated in Fig. 1. It shows that the
potential exhibits three degenerate minima (the so-called vacua of the
model), localized at $\phi^{(0)}=0$ (the central vacuum) and 
$\phi^{(\pm)}=\pm(3/2)^{1/2}\phi_0$.

\begin{figure}[h]
\centering
\includegraphics[width=8.0cm]{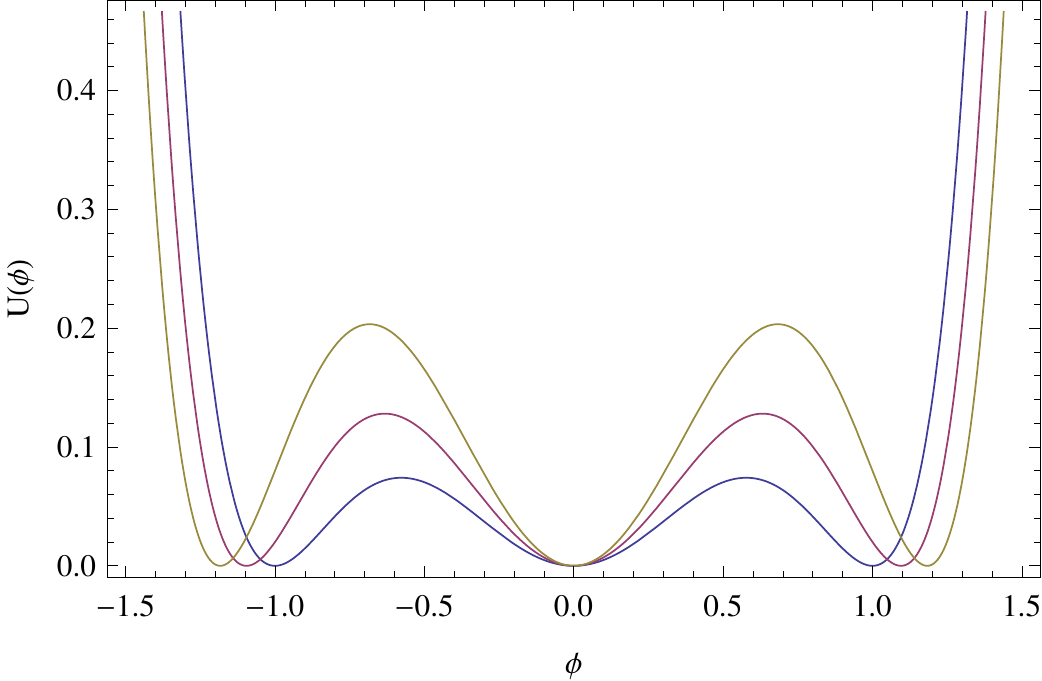}
\caption{{}Profile of the potential (\protect\ref{5}) for some selected
values of $\protect\phi_0$ and $\protect\omega$.}
\label{teste:oscillons}
\end{figure}

In the next Section, we follow the usual approach to derive oscillons in 1+1
space-time dimensions.

\bigskip

\section{Oscillon configurations}

First of all, oscillons are time-periodic structures. Thus, for each fixed
point $(T;X)$, we consider solutions of Eq. (\ref{4}) on the (Euclidean)
ellipse

\begin{equation}
\frac{\tau^2}{T^2}+\frac{\chi^2}{X^2}=1.  \label{6}
\end{equation}

\noindent Next, oscillons are also localised in space. Hence, we regard the
scale mapping $\chi=X\epsilon$, where $\epsilon$ is a small, positive
parameter, $0<\epsilon <<1$. Therefore, Eq. (\ref{6}) shows that the (of
course, positive) time coordinate might transform as $0<\tau=T(1-\epsilon
^{2})^{1/2}$. \textcolor{black}{We notice that those transformations naturally appear in the context of perturbation theory, throught use of so-called multiscale expansion method \cite{multibook}. Within that approach, one can introduce several time and space variables, which may be scaled differently and regarded as independent. Thus, it is possible to identify such scalings from the equations of motion.}

Given the above considerations, Eq. (\ref{4}) becomes

\begin{eqnarray}
&&\omega^2\phi\left(1-\frac{8\phi^2}{3\phi_0^2} 
+\frac{4\phi^4}{3\phi_0^4}\right) 
+(1-\epsilon^2)\partial_{\tau\tau }\phi 
-\epsilon^2\partial_{\chi\chi}\phi  \notag \\
&&-\frac{Hh\epsilon (1-\epsilon ^{2})^{1/2}\phi ^{H-1}\partial _{\tau }\phi
\partial _{\chi} \phi }{(1+2c)^{1/2}(1-2d)^{1/2}}  \notag \\
&&-\frac{2h\epsilon (1-\epsilon ^{2})^{1/2}\phi ^{H}\partial _{\tau\chi}\phi 
} {(1+2c)^{1/2}(1-2d)^{1/2}}=0,  \label{7}
\end{eqnarray}

\noindent where use has been made of Eq. (\ref{5}). Obviously, now 
$\phi=\phi(\tau;\chi)$.

Departing from Eq. (\ref{7}), we can derive small amplitude oscillons which
are localised in the central vacuum of the model described by Eq. (\ref{5}).
The usual approach to accomplish this task in $1+1$ dimensions is to
consider the expansion of the field $\phi$ about $\phi^{(0)}=0$ in a power
series of the small parameter $\epsilon$. \textcolor{black}{However, it is worth to emphasize that, some time ago, a new kind of approach to nonli\-near solutions, which introduces a class of configurations now known as flat-top oscillons, has been developed by Amin and Shirokoff \cite{aminPRD2010}.}

Now, we notice that if $H$ is an
integer which satisfies the condition $H\geq2$, then Eq. (\ref{7}) just exhibits
odd powers of $\phi$. \textcolor{black}{Since only terms up to $\epsilon^{3}$-order will be fundamental to determine the oscillon profile,} we choose $H=2$ for the simplest case,

\begin{eqnarray}
&&\omega^2\phi\left(1-\frac{8\phi^2}{3\phi_0^2} 
+\frac{4\phi^4}{3\phi_0^4}\right)+(1-\epsilon^2)\partial_{\tau\tau }\phi
-\epsilon^2\partial_{\chi\chi}\phi  \notag \\
&&-\frac{2h\epsilon (1-\epsilon ^{2})^{1/2}\phi\partial _{\tau }\phi
\partial _{\chi} \phi } {(1+2c)^{1/2}(1-2d)^{1/2}}  \notag \\
&&-\frac{2h\epsilon (1-\epsilon ^{2})^{1/2}\phi ^{2}\partial _{\tau\chi}\phi 
} {(1+2c)^{1/2}(1-2d)^{1/2}}=0.  \label{8}
\end{eqnarray}

\textcolor{black}{Note that Eq. (\ref{8}) is odd in the field $\phi$. Thus, to obtain oscillon solutions, $\phi$ may be expanded as an asymptotic series in just odd powers of $\epsilon$, namely}

\begin{equation}
\phi=\sum_{n=1}^{\infty }\epsilon ^{2n-1}\phi _{2n-1}\text{.}  \label{9}
\end{equation}

Substituting Eq. (\ref{9}) in Eq. (\ref{8}), the resulting series expansion
yields

\begin{eqnarray}
&&\epsilon(\partial_{\tau\tau}\phi_{1}+\omega^{2}\phi_{1})+  \notag \\
&&\epsilon^{3}\left[(\partial_{\tau\tau}\phi_{3}+
\omega^{2}\phi_{3})-\left(\partial_{\tau\tau}\phi_{1}+
\partial_{\chi\chi}\phi_{1}+ 
\frac{8\omega^2}{3}\frac{\phi_{1}^{3}}{\phi_0^2}
\right)\right]+  \notag \\
&&\mathcal{O}(\epsilon^5)=0.  \label{10}
\end{eqnarray}

\textcolor{black}{From Eq. (\ref{10}), the coefficients of $\epsilon$ and $\epsilon^3$ lead to}

\begin{eqnarray}
\partial _{\tau \tau }\phi _{1}+\omega^{2}\phi _{1} &=&0,  \label{11} \\
\partial _{\tau \tau }\phi _{3}+\omega^{2}\phi _{3} &=& \partial _{\tau \tau
}\phi _{1}+\partial _{\chi \chi }\phi _{1} +\frac{8\omega^2}{3}\frac{\phi
_{1}^{3}}{\phi_0^2},  \label{12}
\end{eqnarray}

\noindent respectively. \textcolor{black}{Now, since the functions $\phi_n$ are even in both $\tau$ and $\chi$, it follows that an also even function $\varphi(\chi)$ exists. The amplitude of the latter shall decay to zero as $\left|\chi\right|\rightarrow\infty$, and} the solution of Eq. (\ref{11}) may be promptly deduced,

\begin{equation}
\phi _{1}(\tau ;\chi )=\varphi(\chi )\cos (\omega \tau ).  \label{13}
\end{equation}

\noindent Substituting Eq. (\ref{13}) in Eq. (\ref{12}), we obtain

\begin{eqnarray}
\partial_{\tau\tau}\phi_{3}&+&\omega^{2}\left[\phi_{3} 
-\frac{2\varphi^{3}}{3\phi_0^2}\cos(3\omega\tau)\right] 
\notag \\
&=&\left[\varphi_{\chi\chi} 
-\omega^{2}\varphi\left(1-\frac{2\varphi^{2}} 
{\phi_0^2}\right)\right]\cos(\omega\tau),  \label{14}
\end{eqnarray}

\noindent where $\varphi_{\chi\chi}\equiv d^2\varphi/d\chi^2$.

It is clear that the general solution of Eq. (\ref{14}) is neither periodic in time nor localised in space. \textcolor{black}{However, the periodicity of $\phi_{3}$ requires that $\partial_{\tau\tau}\phi_{3}+ \omega^{2} \phi_3$ be orthogonal to $\cos(\omega\tau)$, which imposes the condition}

\begin{equation}
\varphi _{\chi\chi}=\omega ^{2}\varphi\left(1-\frac{2\varphi^{2}} 
{\phi_0^2}\right)  \label{15}
\end{equation}

\noindent to be satisfied. As one may easily check, both $\phi_1$ and $\phi_3$ (thus, also $\phi$) qualify to describe oscillon configurations. \textcolor{black}{Up to a translation, Eq. (\ref{15}) exhibits a nontrivial solution, which exactly tends to zero as $\left|\chi\right|\rightarrow\infty$. Thus, after a straightforward calculation, the solution of Eq. (\ref{15}) can be written in the form}

\begin{equation}
\varphi (\chi)=\phi_0\sech(\omega\chi),  \label{16}
\end{equation}

\noindent and the small amplitude oscillon is determined by (see Eqs. 
(\ref{9}) and (\ref{13}))

\begin{equation}
\phi(\tau;\chi)=\epsilon\phi_0\sech(\omega\chi)\cos(\omega\tau) 
+\mathcal{O}(\epsilon^{3}).  \label{17}
\end{equation}

\noindent By plugging back the original variables, we finally define the
oscillon field as

\begin{equation}
\phi^{(\epsilon)}(t;x)\equiv\epsilon\phi_0\sech(\epsilon\omega_d x) 
\cos\left[\left(1- \frac{\epsilon^2}{2}\right)\omega_c t\right] 
+\mathcal{O}(\epsilon^{3}),  \label{18}
\end{equation}

\noindent where use has been made of the binomial approximation 
$(1-\epsilon^2)^{1/2}\approx(1-\epsilon^2/2)$ for $0<\epsilon\ll1$, and we
have defined $\omega_c\equiv\omega(1+2c)^{-1/2}$ and 
$\omega_d\equiv\omega(1-2d)^{-1/2}$.

A typical profile of $\phi^{(\epsilon)}(t;x)$ is illustrated in Fig. 2. It
shows that the fields indeed oscillate about their corresponding effective
mean value.

\begin{figure}[h]
\centering
\includegraphics[width=8.0cm]{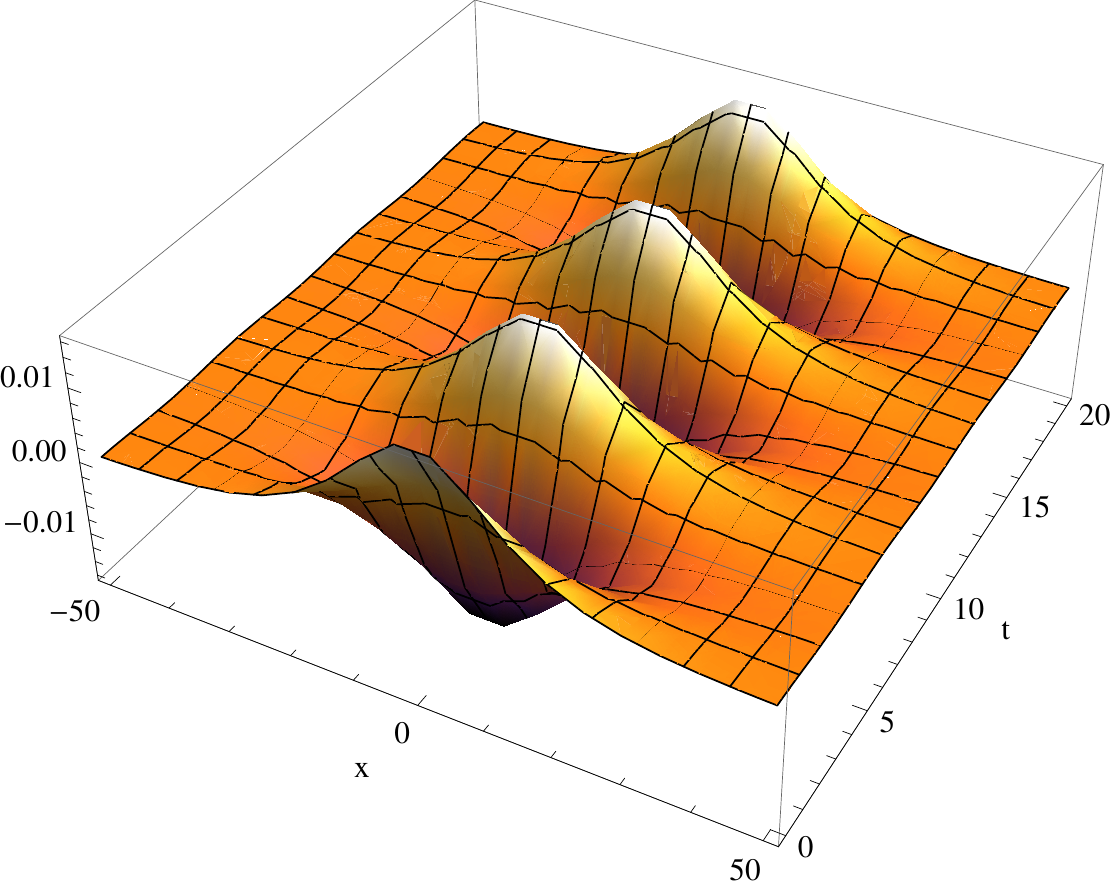}
\caption{{}Typical profile of the oscillon on analysis for 
$\protect\epsilon=0.01$, $\protect\phi_0=(2/3)^{1/2}$, 
$\protect\omega =2^{1/2}$, $c=0$, and $d=0$.}
\label{Fig2:oscillons}
\end{figure}

In Fig. 3, we exhibit the variation of the field energy density with respect
to the potencial coefficients $c$ and $d$. As we have shown, those
quantities imply dilations on the frequency $\omega$.

\begin{figure}[h]
\centering
\includegraphics[width=4.0cm]{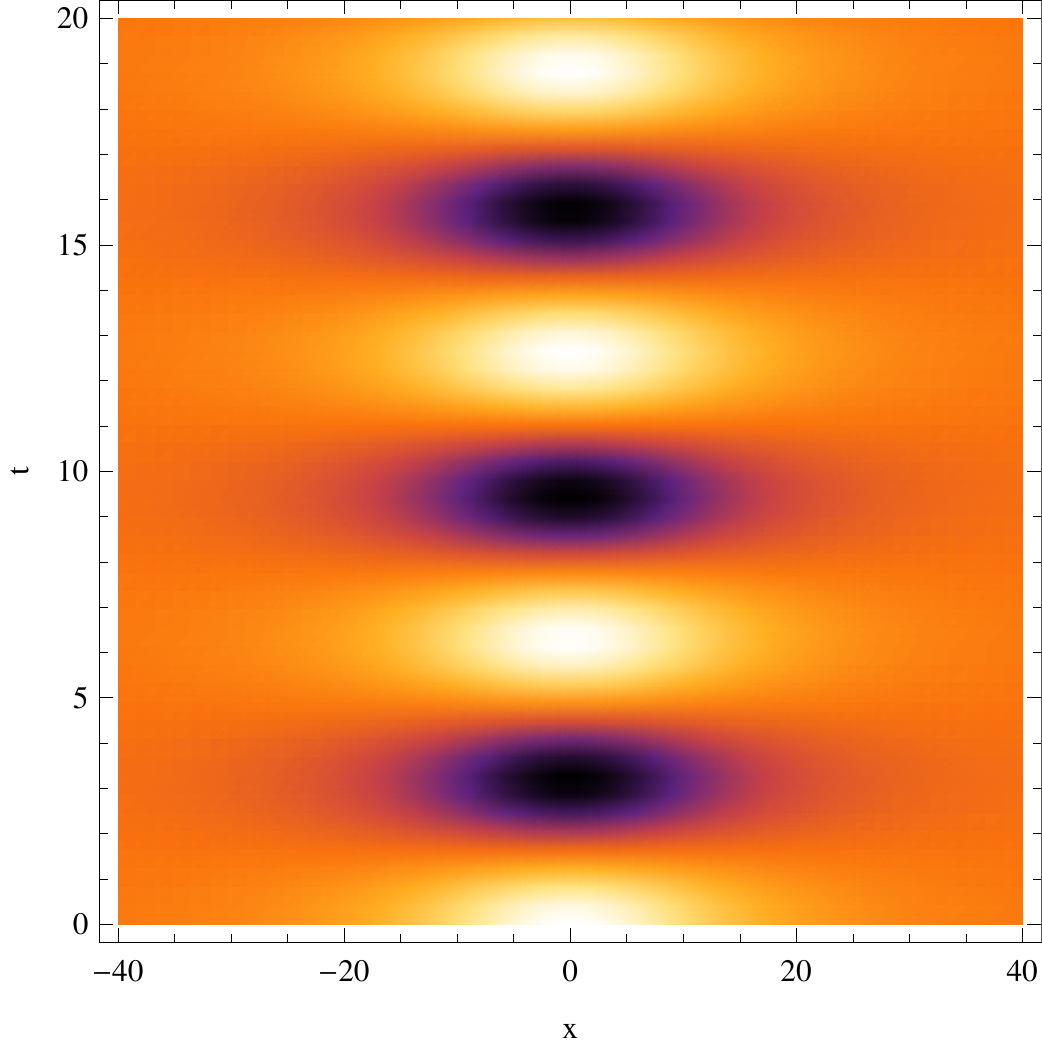}
\includegraphics[width=4.0cm]{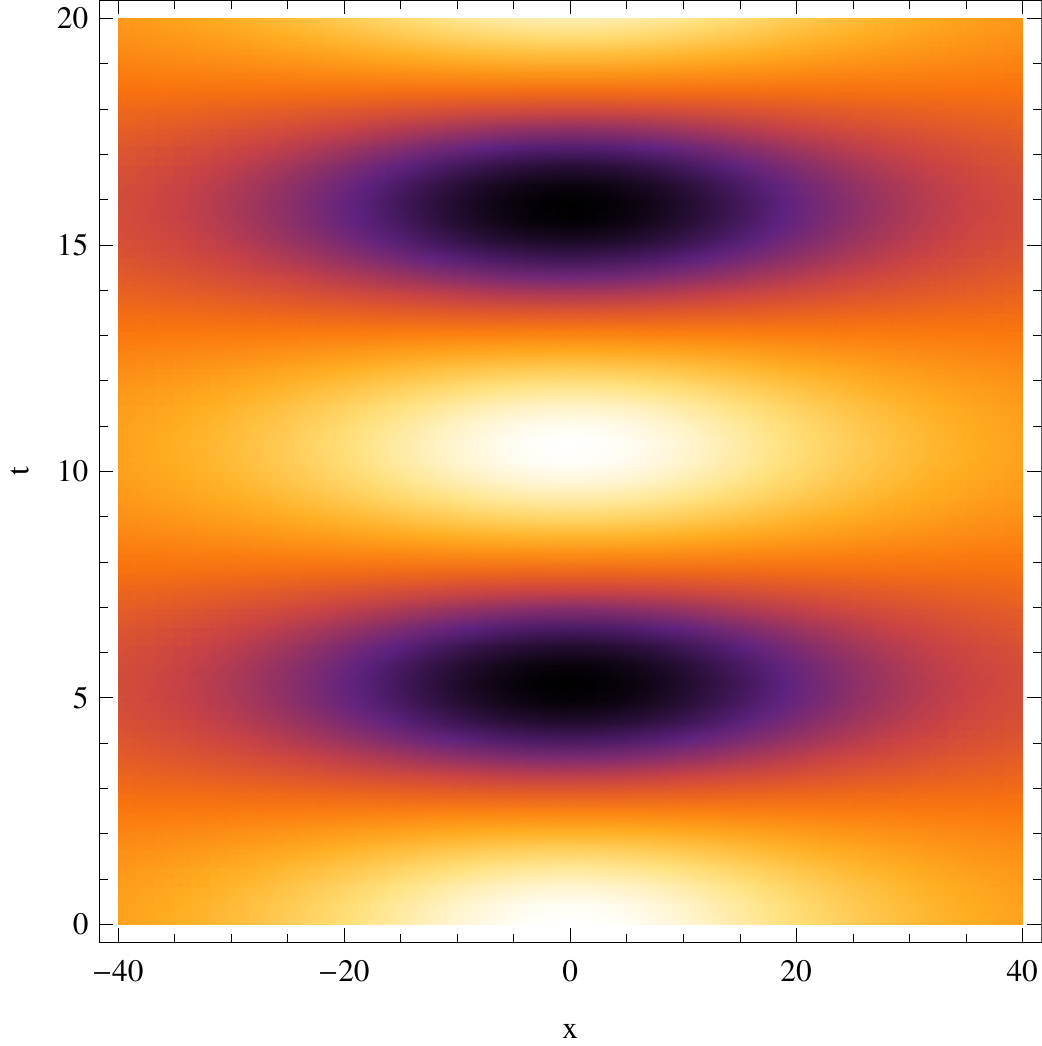}
\caption{{}The variation of the field energy density for 
$\protect\epsilon=0.01$, $\protect\phi_0=(2/3)^{1/2}$, 
$\protect\omega =2^{1/2}$, $a=1$, $b=1$, $f=1$, $g=1$, 
and $h=1$. On the left frame, $c=0$ and $d=0$. On the 
right frame, $c=0.9$ and $d=-1.5$.}
\label{Fig3:energy}
\end{figure}

In the next Section, we discuss the outgoing radiation of the presently 
deduced oscillon configuration.

\section{Outgoing radiation}

In this Section, we follow the steps of Hertzberg \cite{hertzbergPRD2010}, 
who has proposed a method to compute the classical radiation in 1+1 
dimensional Minkowski space-time. The approach assumes that one can write
the solution of the filed equations as 

\begin{equation}
\phi^{(\epsilon)}(t;x)=\phi^{(\epsilon)}_{osc}(t;x)+\zeta (t;x),  
\label{19}
\end{equation}

\noindent where $\phi ^{(\epsilon )}_{osc}(t;x)$ is the oscillon solution and 
$\zeta (t;x)$, a small correction. Thus, by substituting Eq. (\ref{19}) 
into Eq. (\ref{3}), we obtain the linearized formula 

\begin{equation}
\partial _{TT}\zeta-\partial _{XX}\zeta+\omega ^{2}\zeta(T;X) 
=-\Delta(T;X),  \label{20}
\end{equation}

\noindent where use has been made of the space-time dilations and

\begin{equation}
\Delta (T;X)=\partial _{TT}\phi ^{(\epsilon )} 
-\partial _{XX}\phi^{(\epsilon )} 
+\omega ^{2}\phi ^{(\epsilon )}(T;X) 
\label{21}
\end{equation}

\noindent may be interpreted as an external source term, which 
enables us to write the correction field as

\begin{equation}
\zeta (T;X)=-\frac{1}{(2\pi )^{2}}\lim_{\eta\rightarrow 0^{+}}\int d\Omega 
\int dK\frac{\Delta (\Omega ,K)e^{i(KX-\Omega T)}}
{K^{2}-\Omega ^{2}\pm i\eta},
\label{22}
\end{equation}

\noindent with the Fourier component

\begin{equation}
\Delta (\Omega ,K)=\int dT\int dX\Delta(T;X)e^{-i(KX-\Omega T)}.  \label{23}
\end{equation}

Eq. (\ref{23}) shows that the amplitude of the radiation field $\zeta (t;x)$ 
is much smaller than $\epsilon\phi_0$, the oscillon amplitude (see Eq. (\ref{18})). 
This means that the presently given oscillon solutions are actually stable 
configurations.

In the next Section, we establish a possible connection between oscillon
configurations and plasma physics.

\section{Connection with plasma physics}

Ideal magnetohydrodynamics (MHD) is the theory that describes the time and
space evolution of a perfectly conducting fluid subjected to a strong
external magnetic field. Action principles for MHD are usually constructed 
to reflect the interaction of both linear and nonlinear waves with a 
non-uniform background plasma flow \cite{webbPoP2007}. A simple MHD action 
may be written as \cite{ogilvieJPP2016}

\begin{equation}
S_{\mathrm{MHD}}=\int dt \int d^3x \left(\frac{\rho u^2}{2} -\rho\Phi 
-\rho e-\frac{B^2}{2\mu_0}\right),  \label{24}
\end{equation}

\noindent where $\Phi$ is the gravitational potential (for a
non-self-gravitating fluid), $u$ and $B$ are the strengths of the flow and
magnetic fields, respectively, $\rho$ and $e$ are the mass density (per unit
volume) and specific energy (per unit mass), respectively, and $\mu_0$ is
the vacuum magnetic permeability (we use MKS units to comply with the plasma
physics community practice).

An arbitrary variation of the MHD action (\ref{24}) leads to the equation of
motion

\begin{equation}
\rho D_tu_i=-\rho\partial_i\Phi+\partial_jW_{ij},  \label{25}
\end{equation}

\noindent where the differential operator

\begin{equation}
D_t=\partial_t+u_i\partial_i  \label{26}
\end{equation}

\noindent is the material, or convective, derivative (with the abbreviation 
$\partial_i=\partial/\partial x_i$), a repeated index denotes a summation 
over $i=1,2,3$, and the stress tensor (including thermal and magnetic 
pressures) may be read as

\begin{equation}
W_{ij}=-\left(p+\frac{B^2}{2\mu_0}\right)\delta_{ij} +\frac{1}{\mu_0}B_iB_j,
\label{27}
\end{equation}

\noindent with $p=-(\partial e/\partial v)\mid_s$, where $s$ and 
$v=\rho^{-1}$ denote the fluid specific entropy and volume, respectively
(we regard an equation of state in the form $e=e(s,v)$, thus $p$ may be
readily computed from the well-known thermodynamic relation $de=Tds-pdv$,
with $T=(\partial e/\partial s)\mid_v$ denoting the fluid temperature).

Consider now a small perturbation of the (Lagrangean) coordinate $x_i$ 
in the form $x_i\rightarrow x_i+\xi_i$, where we assume that the displacement 
$\xi_i$ satisfies the condition $\mid\xi_i\mid\ll\mid x_i\mid$. Thus, the MHD 
action (\ref{24}) becomes $S_{\mathrm{MHD}}= 
S_{\mathrm{MHD}}(t,x_i;\xi_i,\partial_t\xi_i,\partial_i\xi_j)$, 
thereby yielding the perturbed equation of motion

\begin{equation}
\rho D_{tt}\xi_i=-\rho(\partial_{ij}\Phi)\xi_j+
\partial_j(W_{ijkl}\partial_l\xi_k)+\mathcal{O}(\xi^2),  \label{28}
\end{equation}

\noindent where the differential operator

\begin{equation}
D_{tt}=\partial_{tt}+(\partial_tu_i)\partial_i 
+2u_i\partial_i\partial_t+u_i(\partial_iu_j)\partial_j 
+u_iu_j\partial_{ij}  \label{29}  
\end{equation}

\noindent and the stress tensor

\begin{eqnarray}
W_{ijkl}&=&\left[(\gamma-1)p+\frac{B^2}{2\mu_0}\right] \delta_{ij}%
\delta_{kl} +\left(p+\frac{B^2}{2\mu_0}\right) \delta_{il}\delta_{jk}  
\notag \\
&+&\frac{1}{\mu_0}B_jB_l\delta_{ik} -\frac{1}{\mu_0}(B_iB_j\delta_{kl}
+B_kB_l\delta_{ij}),  \label{30}
\end{eqnarray}

\noindent with $\gamma$ denoting the ratio of specific heats 
at constant pressure to volume (actually, 
$\gamma=(v/p)(\partial^2e/\partial v^2)\mid_s$).

We notice that Eq. (\ref{28}) provides the basis for a nonlinear
perturbation theory for a given ideal MHD flow, since it may be computed at
any order in the Lagrangean displacement $\xi_i$ \cite{ogilvieMNRAS2013}.
With appropriate modifications, the method may be also extended to turbulent
flows \cite{garaudJPP2005}, and even account for certain dissipative effects 
\cite{ogilvieARAA2014}. Further, the proper approach may be modified to
include inertial \cite{lingamPLA} and Hall \cite{charidakosPoP2014} terms,
and yet set forward in the framework of relativistic MHD 
\cite{kawazuraPoP2017}.

Although oscillon configurations have been discussed on 1+1 dimensions in
the previous Sections, and ideal MHD, on 3+1 dimensions in this Section, a
contrast of the relevant equations, (\ref{10}) and (\ref{28}), qualitatively
suggests an analogy between the scalar field $\phi$ and Lagrangean 
displacement $\xi$. Such an observation indicates a perspective for a 
development of the present work. On one hand, we shall extend our oscillon 
solution to 2+1 dimensions, and, on the other, restrict the action principle 
to 2+1 MHD. Thus, we might be able to formulate a quantitative analysis of 
oscillon dynamics in the realm of plasma physics. That will be presented 
in a forthcoming communication.

\section{Conclusion}

In this work, we have reported on the possibility of occurrence of 
extremely long-lived, time-periodic, spatially-localised scalar 
field configurations, the so-called oscillon structures, in the 
fourth state of matter. 

Starting from a scalar theory in 1+1 space-time dimensions, whose 
potential $U$ depends on both the field $\phi$ and its derivatives, 
we have obtained the Euler-Lagrange equations. In particular, the 
self-interacting part of $U$ has been suitably chosen to be a 
$\phi^6$ potential. Small amplitude oscillons have emerged as a 
power series expansion in the scaling parameter $\epsilon$ of the 
solutions of the field equations on an Euclidean ellipse. 
A perturbative analysis has shown that such configurations are 
actually long-lived structures.

A connection with plasma physics has been established in the 
framework of ideal MHD theory in 3+1 dimensions. In that realm, 
the perturbative Lagrangean displacement $\xi$ has been shown 
to play the role of the scalar field $\phi$, as qualitatively 
suggested by the contrast of the relevant equations (\ref{28}) 
and (\ref{10}), respectively.

A quantitative analysis deserves a thorough investigation of 
both formulations by matching them in 2+1 space-time dimensions. 
That will be presented in a forthcoming communication. 

\bigskip

\section*{Acknowledgement}

RACC is partially supported by FAPESP (Foundation for Support to Research of 
the State of S\~ao Paulo) under grants numbers 2016/03276-5 and 2017/26646-5. 
WDP and TF are partially supported by FAPESP, CNPq (National Council for 
Scientific and Technologial Development), and CAPES (Coordination for 
Improvement of University Level Personnel). FEMS is partially supported 
by FAPESP under grant number 2017/20192-2.

\end{document}